\documentclass[a4paper,fleqn]{cas-dc}

\usepackage[numbers]{natbib}

\usepackage{bm}
\usepackage{revsymb}  %---LJW: tmp for \lambdabar
\begin{document}
\let\WriteBookmarks\relax
\def\floatpagepagefraction{1}
\def\textpagefraction{.001}
\shorttitle{}
\shortauthors{}

\title [mode = title]{Calculations of bound-state $\beta^-$-decay half-lives of highly ionized $^{163}$Dy$^{66+}$, $^{187}$Re$^{75+}$ and $^{205}$Tl$^{81+}$ }
%\tnotemark[1,2]

%\tnotetext[1]{This document is the results of the research
%   project funded by the National Science Foundation.}

%\tnotetext[2]{The second title footnote which is a longer text matter
%  to fill through the whole text width and overflow into
%  another line in the footnotes area of the first page.}

%\newcommand{\orcidauthorA}{0000-0002-1128-5860} % Add \orcidA{} behind the author's name
%\newcommand{\orcidauthorB}{0000-0003-0630-5748} % Add \orcidB{} behind the author's name

\author[1]{Yang Xiao}
\address[1]{School of Physical Science and Technology, Southwest University, Chongqing 400715, China}

\author[1]{Long-Jun Wang}
\cormark[1]
\cortext[cor1]{longjun@swu.edu.cn}

%\author[2]{Yang Sun}
%\address[2]{School of Physics and Astronomy, Shanghai Jiao Tong University, Shanghai 200240, China}

%%%%%%%%%%%%%%%%%%%%%%%%%%%%%%%%%%%%%%%%%%%%%%%%%%%%%%%%%%%%
\begin{abstract}
  We propose a theoretical method to calculate the bound-state $\beta$-decay half-lives of highly-ionized atoms, which is based on the combination of the Takahashi-Yokoi model and our recently-developed projected shell model that can take into account both allowed and first-forbidden transitions of nuclear $\beta$ decay. Three examples that are of much experimental interests, $^{163}$Dy$^{66+}$, $^{187}$Re$^{75+}$ and $^{205}$Tl$^{81+}$ are taken for calculations. The ground and low-lying states of related nuclei are described reasonably. The bound-state $\beta$-decay half-lives of $^{163}$Dy$^{66+}$ and $^{187}$Re$^{75+}$ are described within a factor of two and four by our calculations without and with the quenching factors in allowed and first-forbidden transitions. The bound-state $\beta$-decay half-life of the last $s$-process branching point $^{205}$Tl$^{81+}$ is predicted to be 58 and 305 days for cases without and with the quenching factors in calculations. The presented method provides a theoretical way to calculate systematically the bound-state $\beta$-decay half-lives of nuclei from light to heavy ones including odd-mass and even-mass cases for the first time. 
\end{abstract}

%  \includegraphics{figs/grabs.pdf}

%. \item Research highlights item 1
%. \item Research highlights item 2
%. \item Research highlights item 3


\maketitle
%%%%%%%%%%%%%%%%%%%%%%%%%%%%%%%%%%%%%%%%%%%%%%%%%%%%%%%%%%%%%%%%%%%%%%%%%%%%%%%%%%%%%%%%%%%%%%%%%%%%%%%%%%%%%
%\section{Introduction}
%\label{sec:Intro}

Nuclear weak-interaction processes are crucial for nuclear physics, particle physics and nuclear astrophysics. For example, the nuclear $\beta^-$ decay rates are key ingredients for understanding the stellar nucleosynthesis by the slow ($s$-) and rapid ($r$-) neutron-capture processes \cite{s_process_RMP_2011, r_process_RMP_2021}. In the case of a neutral atom, $\beta^-$ decay refers to the decay of a neutron in the nucleus to a proton, emitting an electron and an antineutrino in continuum states, as shown in Eq. (\ref{eq.schematic}a). In stellar environments with high temperature and density, on the other hand, atoms get highly ionized to be hydrogen-like or even bare nuclei, so that another exotic decay channel is possible, that is, the electron is not emitted to the continuum but occupying directly bound states of electron orbitals of daughter atoms \cite{Litvinov_2011_Rep_Prog_Phys}, see Eq. (\ref{eq.schematic}b). The latter (former) is referred to as the bound-state $\beta^-$ decay, $\beta_{\text{b}}$-decay (continuum $\beta^-$ decay, $\beta_{\text{c}}$-decay).
\begin{subequations} \label{eq.schematic}
\begin{eqnarray} 
  ^{A}_{Z}\text{X}_{N}^{0+}  &\rightarrow& ^{\ \ \ A}_{Z+1}\text{Y}_{N-1}^{1+} + e^- + \tilde{\nu}_e , \\
  ^{A}_{Z}\text{X}_{N}^{Z+}  &\rightarrow& ^{\ \ \ A}_{Z+1}\text{Y}_{N-1}^{Z+}       + \tilde{\nu}_e , 
\end{eqnarray}
\end{subequations}

The $\beta_{\text{b}}$-decay concept was first proposed by Daudel \textit{et al.} in 1947 \cite{Daudel_first_1947}. In 1980s, Takahashi and Yokoi presented a method to calculate the $\beta_{\text{b}}$-decay rates of highly ionized heavy atoms \cite{Takahashi_Yokoi_1983_NPA} and predicted that the stellar $\beta^-$-decay rates of many $s$-process nuclei could be enhanced greatly when the $\beta_{\text{b}}$-decay channel is taken into account further \cite{TY_table_1987}. In 1992, the first $\beta_{\text{b}}$-decay was observed by Jung \textit{et al.} for the bare $^{163}$Dy$^{66+}$ nucleus in a heavy-ion storage ring \cite{Jung_163Dy_PRL_1992}. The $^{163}$Dy is a $s$-process branching point, with its measured half-life as 47$^{+5}_{-4}$ days, the $s$-process temperature at this branching point was estimated to be $T=0.34(10)$ GK \cite{Litvinov_2011_Rep_Prog_Phys}. In 1996, Bosch \textit{et al.} observed the $\beta_{\text{b}}$-decay of bare $^{187}$Re$^{75+}$ nucleus \cite{Bosch_187Re_PRL_1996}. The $\beta_{\text{b}}$-decay half-life was measured to be only 33 years, as compared with the very long $\beta_{\text{c}}$-decay half-life of 42 Gyr of neutral $^{187}$Re atoms, leading to the $^{187}$Re/$^{187}$Os pair as a cosmo-thermometer instead of a nuclear cosmo-chrononmeter \cite{Litvinov_EPJA_2023}. In 2005, the $\beta_{\text{b}}$-decay of bare $^{207}$Tl$^{81+}$ was studied \cite{Ohtsubo_207Tl_PRL_2005}, and very recently, the measurement of the $\beta_{\text{b}}$-decay of $^{205}$Tl$^{81+}$ was conducted at GSI and the corresponding half-life will be analyzed soon \cite{Litvinov_EPJ_Web_2023}, which will be very helpful for understanding the termination of the $s$ process as $^{205}$Tl serves as the last $s$-process branching point. 

Theoretically, calculations of the $\beta_{\text{b}}$-decay rates (half-lives) are basically based on the Takahashi-Yokoi model \cite{Takahashi_Yokoi_1983_NPA} where nuclear transition strengths, including both allowed and forbidden transitions, are indispensable key inputs. However, in most of the relevant calculations, unknown nuclear transition strengths are not from sophisticated nuclear many-body calculations, but estimated empirically by similar transitions of neighboring nuclei with available data \cite{TY_table_1987, Gupta_PRC_2019_bound_state_beta, Liu_Shuo_PRC_2021}, which may limit interesting predictions. The $\beta_{\text{b}}$-decay channel is expected to enhance to large extent the stellar decay rates of many $s$-process related nuclei, such as $^{106}$Ru, $^{150}$Nd, $^{157}$Gd, $^{160}$Gd, $^{163}$Dy, $^{171}$Tm, $^{179}$Hf, $^{187}$Re, $^{194}$Os, $^{193}$Ir, $^{195}$Pt, $^{205}$Tl, $^{210}$Pb etc. \cite{TY_table_1987}, which are basically rare-earth nuclei or nuclei near $^{205}$Tl. Therefore, elaborate and practical nuclear many-body models are expected for studying and predicting $\beta_{\text{b}}$-decay rates. 

Over the past decade, the traditional projected shell model (PSM) \cite{PSM_review, Sun_1996_Phys_Rep} was developed to include extended large configuration space taking into account high-order quasiparticle (qp) configurations with the help of the Pfaffian and other algorithms \cite{Mizusaki_2013_PLB, LJWang_2014_PRC_Rapid, LJWang_2016_PRC, ZRChen_2022_PRC, BLWang_2022_PRC}. Interesting problems of nuclear high-spin states had then been studied successfully by the extended PSM later \cite{LJWang_2016_PRC, LJWang_2019_JPG, LJWang_PLB_2020_chaos}. In recently years, the PSM was further developed for describing allowed Gamow-Teller (GT) transition and stellar weak-interaction rates \cite{Z_C_Gao_2006_GT, LJWang_2018_PRC_GT, LJWang_PLB_2020_ec, LJWang_2021_PRL, LJWang_2021_PRC_93Nb, zrchen2023symm, ZRChen_PLB2024} as well as nuclear $\beta$ spectrum \cite{FGao_PRC2023}. Besides, very recently, Wang \emph{et al}., presented a new development of the PSM for description of first-forbidden transition of nuclear $\beta$ decay for the first time \cite{BLWang_1stF_2024}. It is then straightforward to combine the current PSM that can treat both allowed GT and first-forbidden transitions, with the Takahashi-Yokoi model to apply to systematical calculations and predictions of stellar decay rates of the above mentioned rare-earth nuclei or nuclei near $^{205}$Tl \cite{TY_table_1987} and other heavier nuclei as shown in Refs. \cite{Gupta_PRC_2019_bound_state_beta, Liu_Shuo_PRC_2021} for which the $\beta_{\text{b}}$-decay channel plays crucial roles. In this work we first present the calculations of the $\beta_{\text{b}}$-decay rates for typical examples of $^{163}$Dy$^{66+}$, $^{187}$Re$^{75+}$ and $^{205}$Tl$^{81+}$ that are actually of experimental interests, i.e., 
\begin{subequations} \label{eq.examples}
\begin{eqnarray} 
  ^{163}\text{Dy}^{66+} &\rightarrow& ^{163}\text{Ho}^{66+} + \tilde{\nu}_e,  \\
  ^{187}\text{Re}^{75+} &\rightarrow& ^{187}\text{Os}^{75+} + \tilde{\nu}_e,  \\
  ^{205}\text{Tl}^{81+} &\rightarrow& ^{205}\text{Pb}^{81+} + \tilde{\nu}_e,  
\end{eqnarray}
\end{subequations}

The rate of $\beta_{\text{b}}$ decay from highly ionized parent $^{A}_{Z}\text{X}_{N}^{Z+}$ to daughter $^{\ \ \ A}_{Z+1}\text{Y}_{N-1}^{Z+}$ as shown in Eq. (\ref{eq.schematic}b) can be calculated as indicated in Ref. \cite{Takahashi_Yokoi_1983_NPA},
\begin{eqnarray} \label{eq.lambda_b}
  \lambda^{\beta_{\text{b}}}_{\text{I}} = \sum_F \lambda^{\beta_{\text{b}}}_{\text{IF}} 
  = \sum_F \frac{\ln 2}{(ft)_{\text{IF}} } f^\ast_{\text{IF} (m)} \ , 
\end{eqnarray}
with $I$ labeling the initial state (usually ground state or isomer) and the summation over all final ($F$) states within the $Q$-value window. One of the most interesting and crucial properties of $\beta_{\text{b}}$ decay is that its $Q$ value is different from that of $\beta_{\text{c}}$ decay. For fully ionized atom (bare nucleus), one can get \cite{Litvinov_2011_Rep_Prog_Phys},
\begin{eqnarray} \label{eq.Q_value}
  Q_{\beta_{\text{b}}} (K, L, \cdots) = Q_{\beta_{\text{c}}} - \Delta B_{e^-} + B_{e^-}^{K, L, \cdots} , 
\end{eqnarray}
where $Q_{\beta_{\text{c}}}$ is the $\beta_{\text{c}}$-decay $Q$ value of neutral atoms defined as the atomic mass difference of the neutral parent and daughter atoms, $Q_{\beta_{\text{b}}}$ is the $\beta_{\text{b}}$-decay $Q$ value of bare nucleus. $B_{e^-}^{K, L, \cdots}$ is the binding energy of the created electron in the daughter atom depending on which ($K$-, $L$-, or other) shell is occupied. $\Delta B_{e^-} = B_n (Z+1) - B_n (Z)$ is the difference of the sum of all electron binding energies for atoms. It is seen from Eq. (\ref{eq.Q_value}) that when $B_{e^-}^{K, L, \cdots} - \Delta B_{e^-} > |Q_{\beta_{\text{c}}}|$ as usually for the case of $K$-shell occupation (see Table \ref{tab1}), $\beta_{\text{b}}$-decay has higher $Q$ value with $Q_{\beta_{\text{b}}} (K, L, \cdots) > Q_{\beta_{\text{c}}}$. On one hand, the increased $Q_{\beta_{\text{b}}}$ would enable transitions to more low-lying states of the daughter nucleus, which may lead to totally different $\beta_{\text{b}}$-decay rate compared with $\beta_{\text{c}}$-decay rate, as for $^{187}$Re in Table \ref{tab1}. On the other hand, stable atom with negative $Q_{\beta_{\text{c}}}$ would become unstable nucleus with positive $Q_{\beta_{\text{b}}}$ as for the two $s$-process branching points $^{163}$Dy and $^{205}$Tl (see Table \ref{tab1}). Actually, even for some nuclei with negative $Q_{\beta_{\text{b}}}$, thermal population of low-lying states of parent nuclei in high-temperature stellar conditions would lead to opening of $\beta_{\text{b}}$-decay channel.

In Eq. (\ref{eq.lambda_b}), $f^\ast$ is the lepton phase volume function in the Takahashi-Yokoi model \cite{Takahashi_Yokoi_1983_NPA},
\begin{eqnarray} \label{eq.f_star}
  f^\ast_{\text{IF} (m)} = \sum_x \sigma_x (\pi / 2) \left[ g_x \text{ or } f_x \right]^2 q^2 \mathcal S_{(m)x},
\end{eqnarray}
with $m=$ a, nu, u labeling allowed, non-unique first-forbidden and unique first-forbidden transitions, with selection rule for spin-parity difference as $\Delta J^{\Delta \pi} = 0^+, 1^+$, $=0^-, 1^-$ and $2^-$ respectively. $\sigma_x$ describes the vacancy of the electron orbital $x$ which lies between zero and unity (unity is adopted here). $\left[ g_x \text{ or } f_x \right]$ is understood as the larger component of the (dimensionless) electron radial wave functions for orbital $x$ evaluated at the nuclear radius $R$, i.e., 
\begin{eqnarray}
  f_x = \frac{P(R)}{R} \lambdabar^{3/2}, \qquad g_x = \frac{Q(R)}{R}\lambdabar^{3/2} ,
\end{eqnarray}
where $\lambdabar = \hbar/m_e c$ is the reduced Compton wavelength of electron, $P$ and $Q$ are the upper- and lower-component radial functions of the electron Dirac wave function in the Coulomb field of the daughter atom as,
\begin{eqnarray}
  \psi_{E\kappa m}(\bm r) = \frac{1}{r} 
  \left( \begin{array}{c}
     P_{E\kappa}(r) \Omega_{\kappa,m}(\hat{\bm r}) \\
    iQ_{E\kappa}(r) \Omega_{-\kappa,m}(\hat{\bm r})
  \end{array} \right),
\end{eqnarray}
where $\Omega(\hat{\bm r})$ labels the corresponding spherical spinor, $E$ is the energy of electron and the quantum number $\kappa$ reads as,
\begin{eqnarray}
  \kappa = (l-j)(2j+1), 
\end{eqnarray}
with 
\begin{eqnarray}
  j=|\kappa|-\frac{1}{2}, \qquad 
  l=\left\{ \begin{array}{ll}
  |\kappa|-1 & \text{if } \kappa<0, \\
  |\kappa|   & \text{if } \kappa>0. \\
  \end{array}  \right.
\end{eqnarray}
where $l, j, m$ are the quantum numbers of orbital angular momentum, non-relativistic total angular momentum and its projection respectively.

%----------------------------------------------------------- Table 1: 
\begin{table}[width=1.0\linewidth,cols=5,pos=t]
  \caption{ The $Q$ values between ground states (g.s.) for $\beta_{\text{c}}$ decay of atoms and $\beta_{\text{b}}$ decay of nuclei, where atomic masses and electron binding energies are taken from Refs. \cite{AME2020_CPC_2021, Atom_binding_data} respectively. See the text for details. } \label{tab1}
\begin{tabular*}{\tblwidth}{@{} CCCCCCC@{} }
  \toprule
  Decays & $Q_{\beta_{\text{c}}}$ & $\Delta B_{e^-}$ & $B_{e^-}^{K}$ & $Q_{\beta_{\text{b}}} (K)$ & $B_{e^-}^{L}$ & $Q_{\beta_{\text{b}}} (L)$  \\ 
  (g.s.$\rightarrow$g.s.) &  (keV)  &  (keV)  &  (keV)  &  (keV)  &  (keV)  &  (keV)   \\
  \midrule
  $^{163}$Dy $\rightarrow$ $^{163}$Ho  &  -2.9 & 12.5 &  65.137 & 49.737 & 15.746 &   0.346  \\
  $^{187}$Re $\rightarrow$ $^{187}$Os  &   2.5 & 15.2 &  85.614 & 72.914 & 20.921 &   8.221  \\
  $^{205}$Tl $\rightarrow$ $^{205}$Pb  & -50.6 & 17.3 & 101.336 & 33.436 & 24.938 & -42.962  \\
  \bottomrule
\end{tabular*}
\end{table}

For bound states the following normalization holds, 
\begin{eqnarray}
  \int_0^\infty \left[ P^2(r) + Q^2(r) \right] dr = 1.
\end{eqnarray}
and in this work the $P(r)$ and $Q(r)$ functions for different electron orbital $x$ in Eq. (\ref{eq.f_star}) are calculated numerically by the recent \texttt{RADIAL} subroutine \cite{Radial_code_CPC_2019}. 

In Eq. (\ref{eq.f_star}) $q= Q_{\beta_{\text{b}}} / m_e c^2$ is the dimensionless $Q$ value, and the spectral shape factors $\mathcal S_{(m)x}$ to the lowest order are given by \cite{Takahashi_Yokoi_1983_NPA}, 
\begin{eqnarray} \label{Smx}  
  \mathcal S_{(m)x} = 
  \left\{ \begin{array}{ll} 
      1             & \text{for } m= \text{a, nu and } x=ns_{1/2}, np_{1/2}, \\ %-1
      q^2           & \text{for } m= \text{u and } x=ns_{1/2}, np_{1/2}, \\ %-2
      \frac{9}{R^2} & \text{for } m= \text{u and } x=np_{3/2}, nd_{3/2}, \\ %-3
      0             & \text{otherwise}.  \\ %-4
  \end{array} \right. 
\end{eqnarray}

In the Takahashi-Yokoi model \cite{Takahashi_Yokoi_1983_NPA}, the comparative half-life, $(ft)_{\text{IF}}$ in Eq. (\ref{eq.lambda_b}), corresponds to the terrestrial value. The partial half-life, $t$, can be calculated by \cite{BLWang_1stF_2024, Zhi_FF_PRC_2013},
\begin{eqnarray} \label{eq.t_here}
  \frac{K_0}{t} 
  = \int_{1}^{Q_{\text{IF}}} C(W) F_0(Z+1, W) pW (Q_{\text{IF}} - W)^{2} dW,
\end{eqnarray}
where $K_0 = 6144 \pm 2 \ s$ \cite{Hardy_2009_PRC}, $W$ and $p = \sqrt{W^2 - 1}$ label the dimensionless total energy and momentum of the electron in units of $m_e c^2$ and $m_e c$ respectively. $F_0$ is the Fermi function, and $Q_{\text{IF}}$ denotes the actual (dimensionless) $Q$ value of the specific transition as, 
\begin{eqnarray} \label{eq.Qif}
  Q_{\text{IF}} = \frac{1}{m_e c^2 }(M_p - M_d + E_I -E_F ), 
\end{eqnarray}
with $E_I (E_F)$ being the nuclear excitation energy of initial (final) state for parent (daughter) nucleus with nuclear mass $M_p (M_d)$. In Eq. (\ref{eq.t_here}) $C(W)$ is the shape factor for the transition, and the phase-space integral $f$ in Eq. (\ref{eq.lambda_b}) corresponds to the integral in Eq. (\ref{eq.t_here}) without the $C(W)$. 

%------------------ For allowed case:

For allowed transitions with $m=$ a, $C(W)$ has no energy dependence \cite{Zhi_FF_PRC_2013}, and is dominated by the Gamow-Teller (GT) transition \cite{Cole_2012_PRC, Sarriguren_2013_PRC, Martinez_Pinedo_2014_PRC},
%\begin{small}
\begin{eqnarray} \label{eq.BGT_if}
  C(W) = B(\text{GT}^-)_{\text{IF}} =
  \Big(\frac{g_A}{g_V} \Big)^2_{\text{eff}} 
  \frac{\big\langle \Psi _{J_F}^{n_{F}} \big\| \sum_{k} \hat{\bm\sigma}^k \hat\tau_{-}^k \big\| \Psi_{J_I}^{n_I} \big\rangle^2 }{2J_{I}+1} , 
\end{eqnarray}
%\end{small}
where $\hat{\bm\sigma}$ ($\hat\tau_{-}$) is the Pauli spin (isospin lowering) operator, the summation runs over all nucleons, and $\Psi_{J}^{n}$ represents the $n$-th eigen nuclear many-body wave function for angular momentum $J$. $(g_{A}/g_{V})_{\text{eff}}$ is the effective ratio of axial and vector coupling constants with corresponding quenching (of the transition operator and/or wave function) for the GT matrix element \cite{A.brown1985, martinez1996, Javier2011PRL, LJWang_current_2018_Rapid, Gysbers_2019_Nat_Phys}, 
\begin{eqnarray} \label{eq.quench_GT}
  \left( \frac{g_{A}}{g_{V}}\right)_{\text{eff}} = f_q(\text{GT}) \left(\frac{g_{A}}{g_{V}} \right)_{\text{bare}} ,
\end{eqnarray}
where $(g_A/g_V)_{\text{bare}} = -1.27641(45)$ \cite{arkisch2019}, and $f_q(\text{GT})$ is the quenching factor for GT transition. Eqs. (\ref{eq.t_here}, \ref{eq.BGT_if}) indicate that for allowed GT transition, $(ft)_{\text{IF}}=K_0 / B(\text{GT}^-)_{\text{IF}}$.

%-------------------- Now for 1st forbidden case:

For first-forbidden transitions, $C(W)$ has explicit energy dependence, which is approximated as \cite{Weidenmuller_FF_RMP_1961, Zhi_FF_PRC_2013, Mougeot_PRC_2015}
\begin{eqnarray} \label{eq.CW}
  C(W) = k(1 + aW + b/W + cW^2) ,
\end{eqnarray}
for non-unique transitions, and \cite{Mougeot_PRC_2015}
\begin{eqnarray} \label{eq.CW2}
  C(W) = k(1 + aW + b/W + cW^2) \left[ q^2 + \lambda_2 p^2 \right] ,
\end{eqnarray}
for unique transitions, where $q$ is the energy of the neutrino and $\lambda_2 = \frac{F_1(Z, W)}{F_0(Z, W)}$ with $F_1$ being the generalized Fermi function \cite{Suhonen_PRC_2017_general_Fermi_for_unique}. The coefficients $k, a, b$ and $c$ depend on the following reduced nuclear matrix elements for first-forbidden transition (see Refs. \cite{Zhi_FF_PRC_2013, BLWang_1stF_2024} for details), 
\begin{subequations} \label{eq.all_ME}
\begin{eqnarray}
  w  &=& -g_A \sqrt{3} \frac{\left\langle \Psi^{n_F}_{J_F} \left\| \sum_k r_k [\bm C^k_1 \otimes \hat{\bm\sigma}^k]^0 \hat{\tau}^k_- \right\| \Psi^{n_I}_{J_I} \right\rangle}{\sqrt{2J_I+1}} , \\
  x  &=& - \frac{\left\langle \Psi^{n_F}_{J_F} \left \| \sum_k r_k \bm C^k_1 \hat{\tau}^k_- \right \| \Psi^{n_I}_{J_I} \right \rangle}{\sqrt{2J_I+1}} , \\
  u  &=& -g_A \sqrt{2} \frac{\left\langle \Psi^{n_F}_{J_F} \left \| \sum_k r_k [\bm C^k_1 \otimes \hat{\bm\sigma}^k]^1 \hat\tau^k_- \right \| \Psi^{n_I}_{J_I} \right \rangle}{\sqrt{2J_I+1}} , \\
  z  &=& 2g_A \frac{\left\langle \Psi^{n_F}_{J_F} \left \| \sum_k r_k [\bm C^k_1 \otimes \hat{\bm\sigma}^k]^2 \hat\tau^k_- \right \| \Psi^{n_I}_{J_I} \right \rangle}{\sqrt{2J_I+1}} , \\ 
  w' &=& -g_A \sqrt{3} \frac{\left\langle \Psi^{n_F}_{J_F} \left \| \sum_k \frac{2}{3}r_k I(r_k) [\bm C^k_1 \otimes \hat{\bm\sigma}^k]^0 \hat\tau^k_- \right \| \Psi^{n_I}_{J_I} \right \rangle}{\sqrt{2J_I+1}} , \nonumber \\ \\
  x' &=& - \frac{\left\langle \Psi^{n_F}_{J_F} \left \| \sum_k \frac{2}{3}r_k I(r_k)  \bm C^k_1 \hat\tau^k_- \right \| \Psi^{n_I}_{J_I} \right \rangle}{\sqrt{2J_I+1}} , \\
  u' &=& -g_A \sqrt{2} \frac{\left\langle \Psi^{n_F}_{J_F} \left \| \sum_k  \frac{2}{3}r_k I(r_k) [\bm C^k_1 \otimes \hat{\bm\sigma}^k]^1 \hat\tau^k_- \right \| \Psi^{n_I}_{J_I} \right \rangle}{\sqrt{2J_I+1}} , \nonumber \\ \\
  \xi'\nu &=& \frac{g_A\sqrt{3}}{M_0} \frac{\left\langle \Psi^{n_F}_{J_F} \left \| \sum_k  [\hat{\bm\sigma}_k \otimes \bm\nabla^k]^0 \hat\tau^k_- \right \| \Psi^{n_I}_{J_I} \right \rangle}{\sqrt{2J_I+1}} , \\
  \xi' y  &=& - \frac{1}{M_0} \frac{\left\langle \Psi^{n_F}_{J_F} \big\| \sum_k  \bm\nabla^k \hat\tau^k_- \big\| \Psi^{n_I}_{J_I} \right \rangle}{\sqrt{2J_I+1}}  .
\end{eqnarray} 
\end{subequations}
where $\bm C_{lm}$ is proportion to the spherical harmonics, and $I(r)$ is the radial function that is related to nuclear charge distribution \cite{Zhi_FF_PRC_2013, BLWang_1stF_2024}. 

As in the allowed GT transition of $\beta$ decay \cite{A.brown1985, martinez1996} and in the double $\beta$ decay \cite{Javier2011PRL, LJWang_current_2018_Rapid}, quenching factors for nuclear matrix elements (transition operators and/or wave functions) of first-forbidden transitions in Eq. (\ref{eq.all_ME}) may be introduced as well \cite{Zhi_FF_PRC_2013}. In Ref. \cite{Zhi_FF_PRC_2013} the following quenching factors are adopted,
\begin{align} \label{eq.quench}
  f_q(\xi'\nu) &= 1.266,          \hspace{3.95em} f_q(w) = f_q(w') = 0.66, \nonumber \\
  f_q(x)       &= f_q(x') = 0.51, \quad f_q(u) = f_q(u') = 0.38, \nonumber \\
  f_q(z)       &= 0.42. 
\end{align}

It is seen that to calculate and predict the $\beta_{\text{b}}$-decay rates and half-lives, the key nuclear inputs are the different reduced nuclear matrix elements in Eqs. (\ref{eq.BGT_if}, \ref{eq.all_ME}), where nuclear many-body wave function $\Psi_{J}^{n}$ should be written in the laboratory frame with good angular momentum and parity as the transitions have strong selection rules. In recent years the PSM was developed by Wang \emph{et al}. to calculate these reduced nuclear matrix elements for both allowed and first-forbidden transitions with large model and configuration spaces \cite{LJWang_2018_PRC_GT, BLWang_1stF_2024}, where exact angular-momentum projection is adopted, i.e., 
\begin{eqnarray} \label{eq.wave_function}
  | \Psi^{n}_{JM} \rangle = \sum_{K\kappa} f_{K\kappa}^{Jn} \hat{P}_{MK}^{J} | \Phi_{\kappa} \rangle ,
\end{eqnarray}
where $\Phi_{\kappa}$ includes the many-body qp vacuum and various qp excitation configurations in the intrinsic frame \cite{LJWang_2014_PRC_Rapid, LJWang_2018_PRC_GT} and the angular-momentum-projection operator reads as,
\begin{eqnarray} \label{AMP_operator}
    \hat{P}^{J}_{MK} = \frac{2J + 1}{8\pi^2} \int d\Omega D^{J\ast}_{MK} (\Omega) \hat{R} (\Omega) ,
\end{eqnarray}

%-----------------------Figure 1:
\begin{figure}
\begin{center}
  \includegraphics[width=0.47\textwidth]{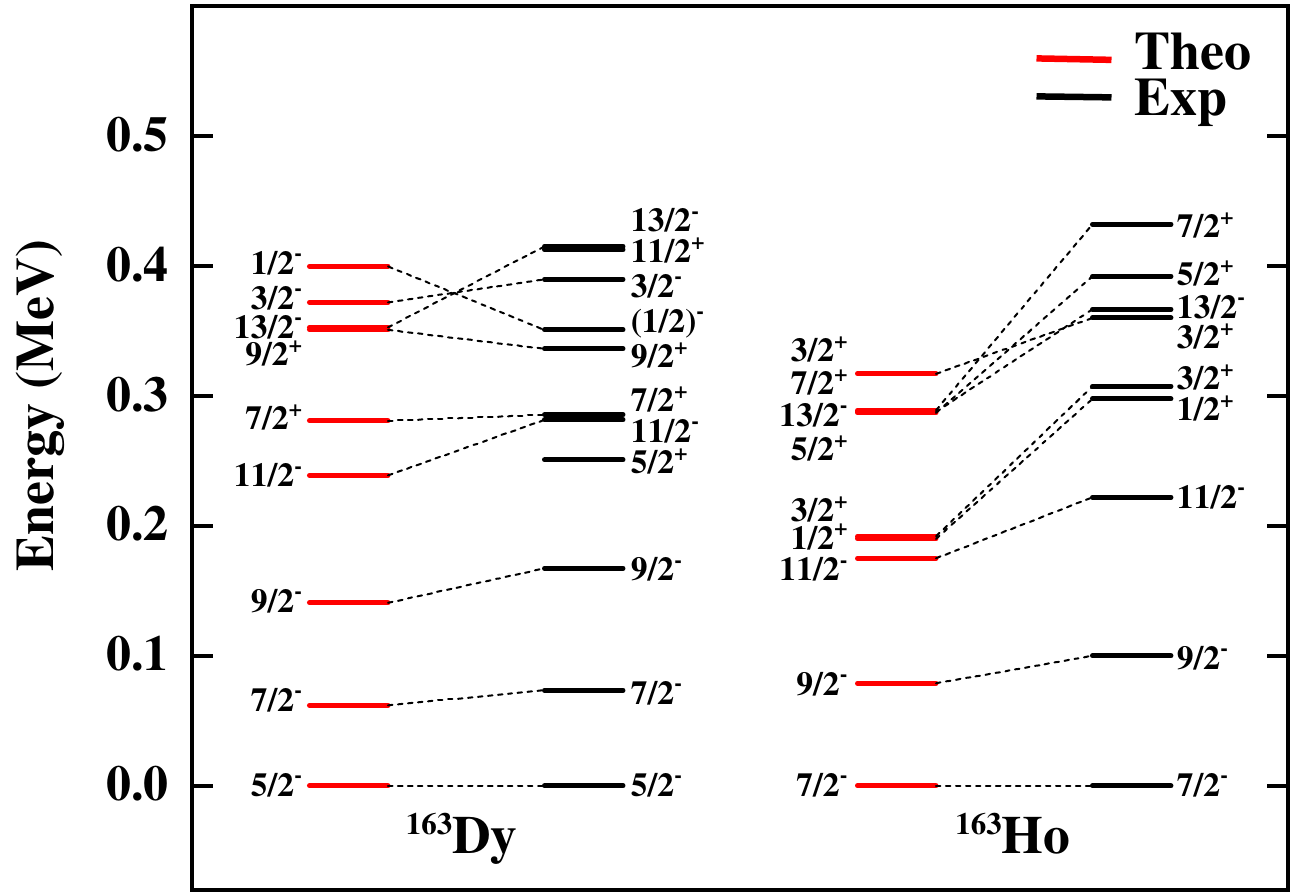}
  \caption{\label{fig:fig1} The calculated excitation energies for low-lying states of $^{163}$Dy and $^{163}$Ho as compared with the data \cite{NNDC}. }
\end{center}
\end{figure}

%-----------------------Figure 2:
\begin{figure}
\begin{center}
  \includegraphics[width=0.45\textwidth]{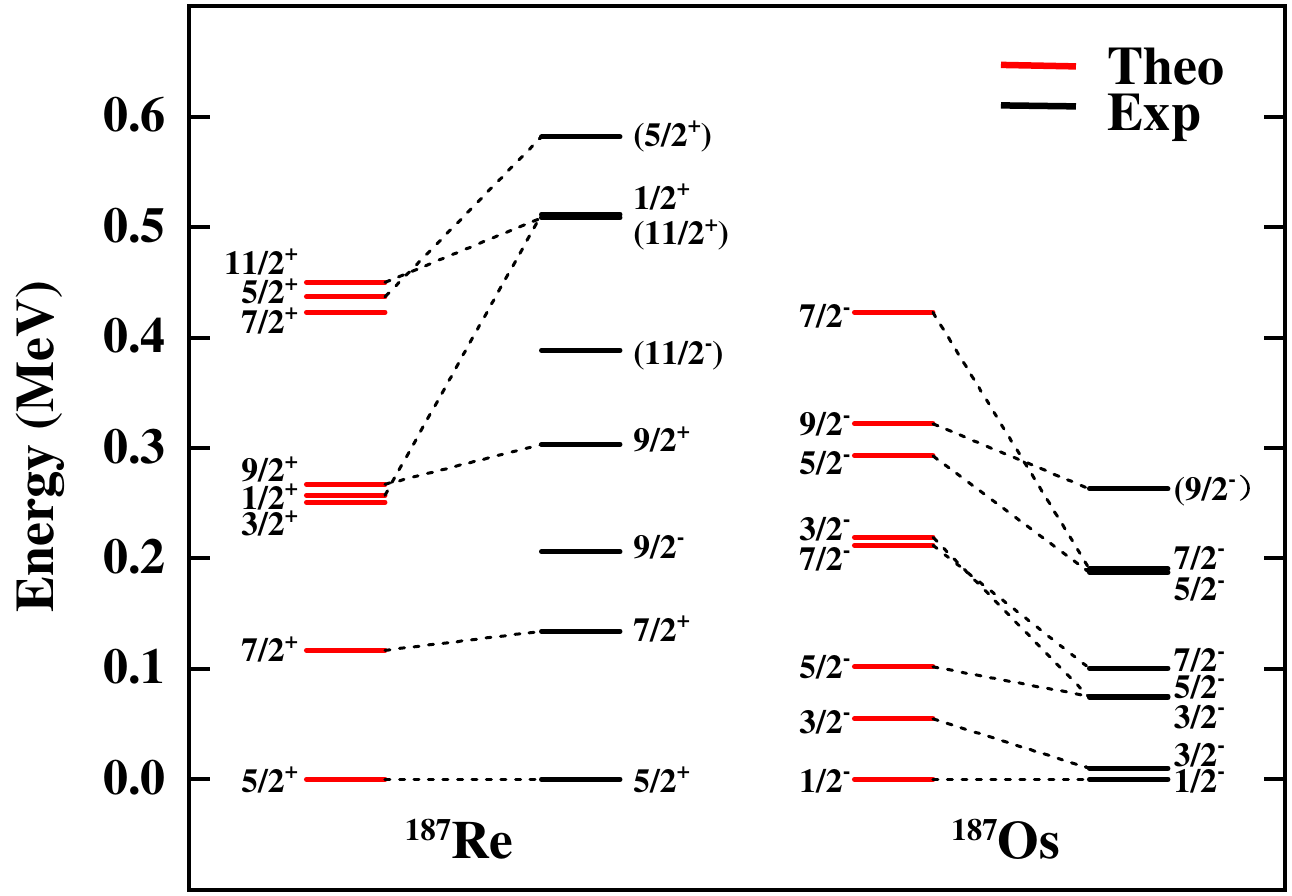}
  \caption{\label{fig:fig2} Same as Fig. \ref{fig:fig1} but for $^{187}$Re and $^{187}$Os. }
\end{center}
\end{figure}

%-----------------------Figure 3:
\begin{figure}
\begin{center}
  \includegraphics[width=0.45\textwidth]{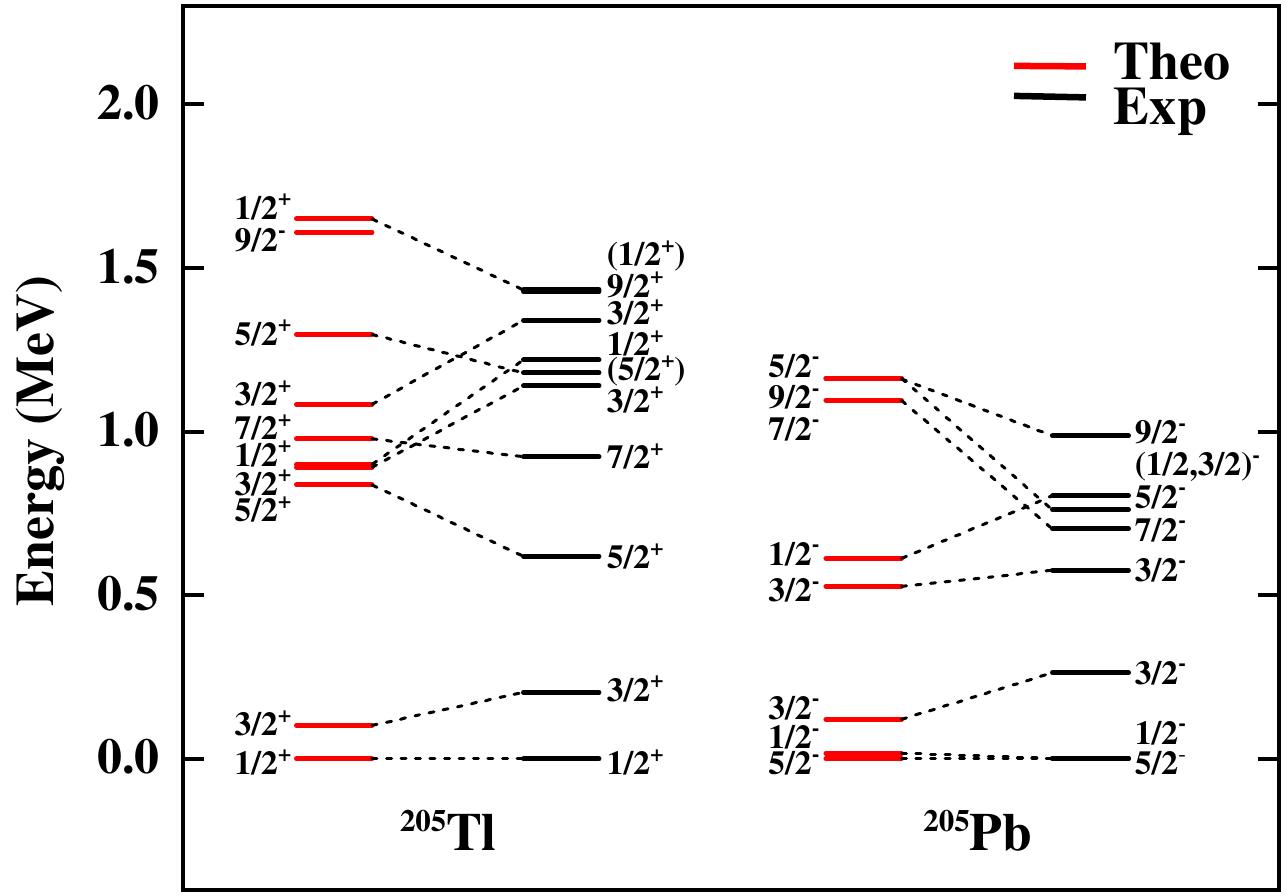}
  \caption{\label{fig:fig3} Same as Fig. \ref{fig:fig1} but for $^{205}$Tl and $^{205}$Pb. }
\end{center}
\end{figure}

%-----------------------Figure 4:
\begin{figure}
\begin{center}
  \includegraphics[width=0.402\textwidth]{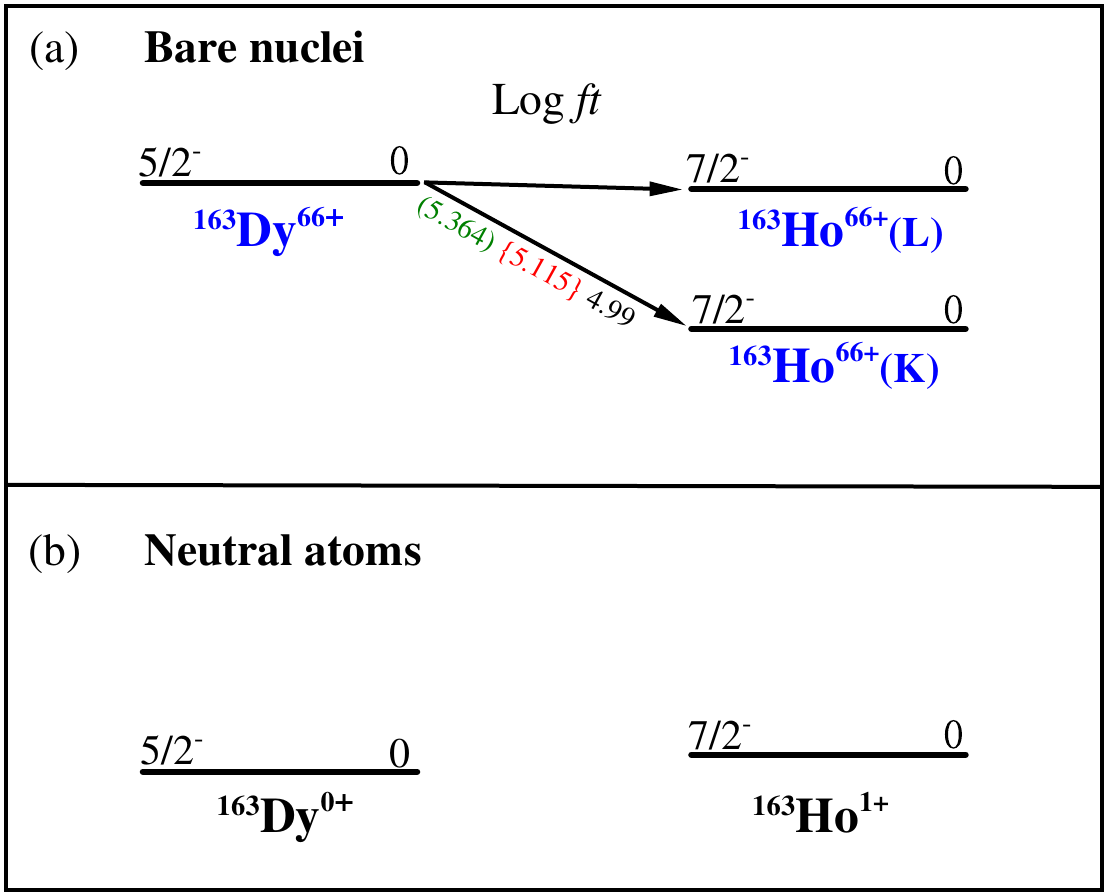}
  \caption{\label{fig:fig4} The decay scheme for neutral and fully ionized $^{163}$Dy with calculated Log$ft$ with and without quenching factors. }
\end{center}
\end{figure}

%-----------------------Figure 5:
\begin{figure}
\begin{center}
  \includegraphics[width=0.387\textwidth]{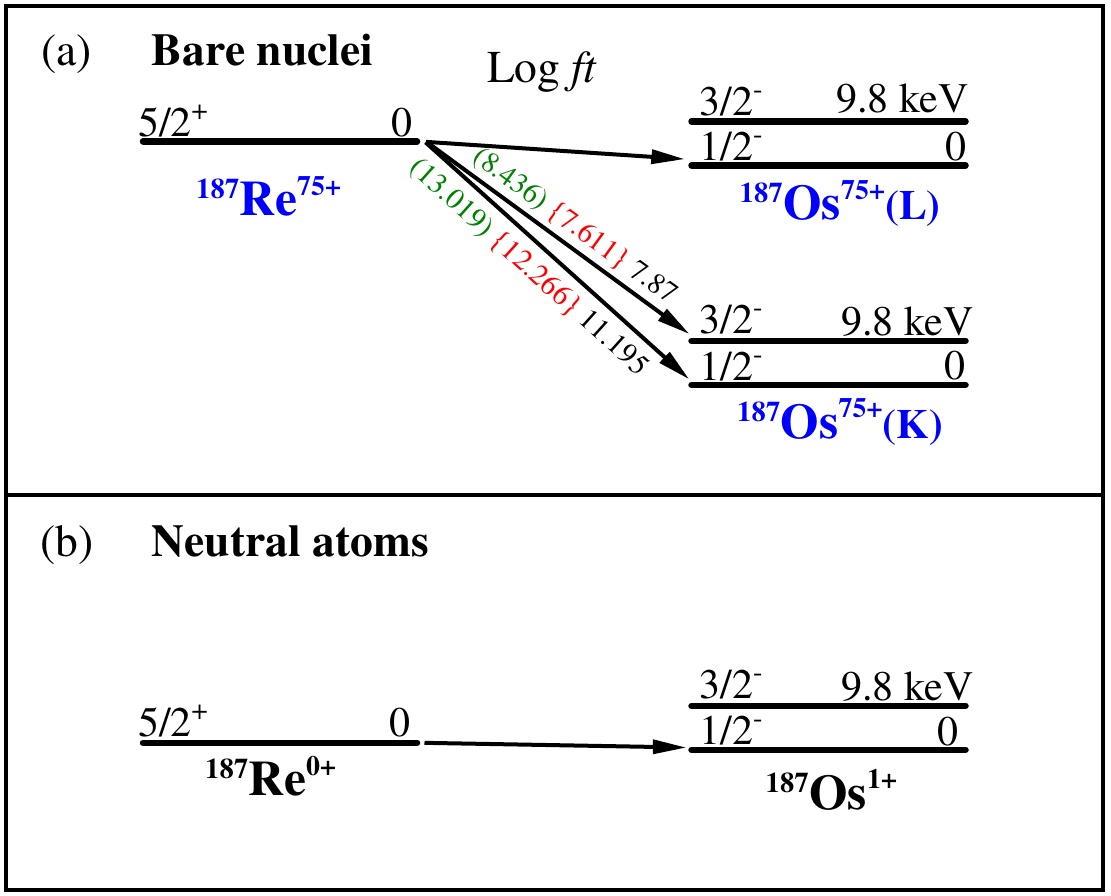}
  \caption{\label{fig:fig5} Same as Fig. \ref{fig:fig4} but for $^{187}$Re. }
\end{center}
\end{figure}

%-----------------------Figure 6:
\begin{figure}
\begin{center}
  \includegraphics[width=0.388\textwidth]{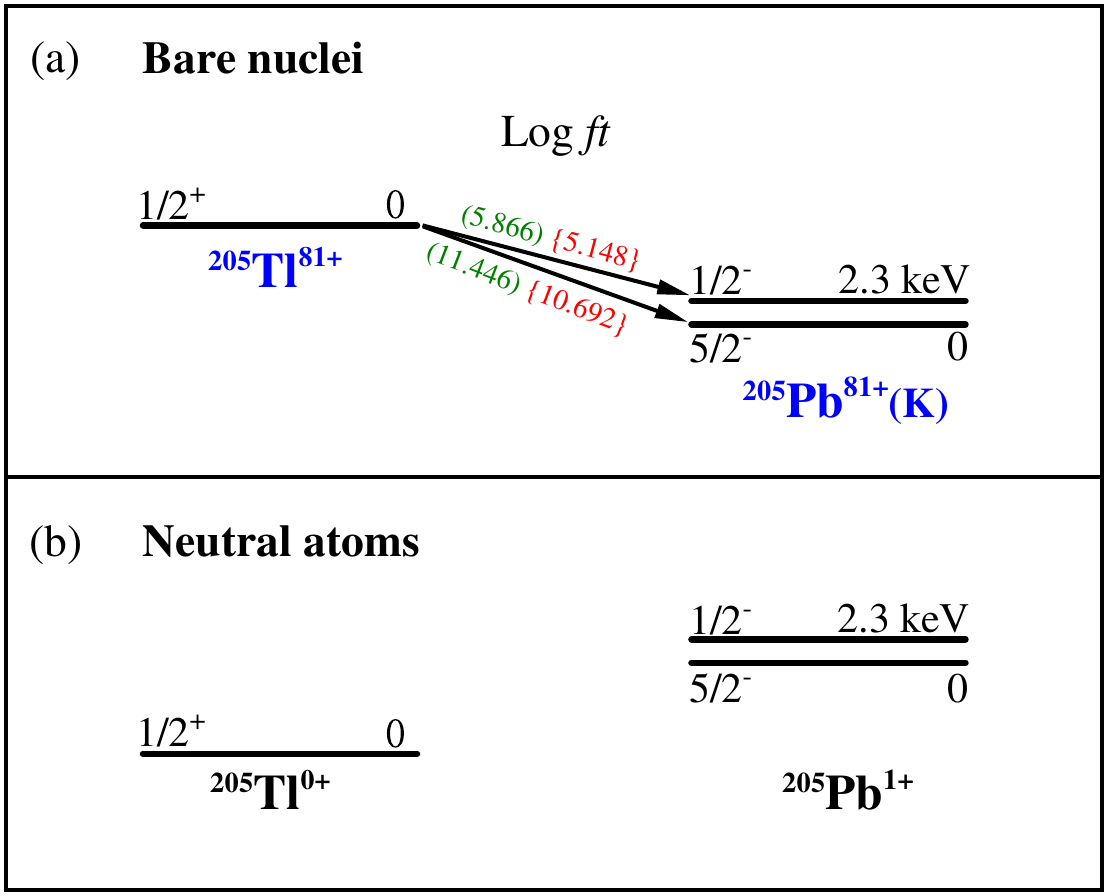}
  \caption{\label{fig:fig6} Same as Fig. \ref{fig:fig4} but for $^{205}$Tl. }
\end{center}
\end{figure}

where $D^J_{MK}$ ($\hat R$) is the Wigner $D$ function (rotation operator) with respect to the Euler angle $\Omega$ \cite{ZRChen_2022_PRC, BLWang_2022_PRC} with $M$ ($K$) being the spin projection in the laboratory (intrinsic) frame. $f$ in Eq. (\ref{eq.wave_function}) labels the expansion coefficients that can be obtained by solving the corresponding eigen equation. The projection operator transforms the description of nuclei from the intrinsic to the laboratory frame. In this work we combine the current PSM that can treat both allowed GT \cite{LJWang_2018_PRC_GT, ZRChen_PLB2024} and first-forbidden transitions \cite{BLWang_1stF_2024}, with the Takahashi-Yokoi model so that the $\beta_{\text{b}}$-decay rates and half-lives of candidate nuclei, especially for the rare-earth nuclei and heavier nuclei, can be calculated and predicted by elaborate nuclear many-body method.

%----------------------------------------------------------- Table 2: 
\begin{table*}[width=2.0\linewidth,cols=10,pos=t]
  \caption{The calculated Log$ft$ and half-lives for bound-state $\beta$ decays of $^{163}$Dy$^{66+}$, $^{187}$Re$^{75+}$ and $^{205}$Tl$^{81+}$ with and without the quenching factors in allowed and first-forbidden transitions as compared with available data \cite{Jung_163Dy_PRL_1992, Bosch_187Re_PRL_1996}.  } 
  \label{tab2}
\begin{tabular*}{\tblwidth}{@{} CCCCCCCCCC@{} }
  \toprule
  Nuclei & Transition & Type & \multicolumn{2}{c}{Exp.} & $f^\ast_{\text{IF}}$ & \multicolumn{4}{c}{Theo.(PSM)} \\ \cline{4-5} \cline{7-10}
         & ($J^\pi_I \rightarrow J^\pi_F$) &  & Log$ft$ & $T_{1/2}(\beta^-_{\text{b}})$ &  & 
         Log$ft$(w/o) & $T_{1/2}(\beta^-_{\text{b}})$ (w/o) & Log$ft$(with) & $T_{1/2}(\beta^-_{\text{b}})$ (with)   \\
  \midrule
  $^{163}$Dy $\rightarrow$ $^{163}$Ho & $5/2^- \rightarrow 7/2^-$ & a & 4.99 & 47$^{+5}_{-4}$ d \cite{Jung_163Dy_PRL_1992} & 0.02266428 
  & 5.115 & 66.54942 d & 5.364 & 118.07127 d  \\ \hline  %---1st
  \multirow{2}{*}{ $^{187}$Re $\rightarrow$ $^{187}$Os }  & 
    $5/2^+ \rightarrow 1/2^-$ &  u & 11.195  & \multirow{2}{*}{32.9$^{+2}_{-2}$ y \cite{Bosch_187Re_PRL_1996} } & 0.00111926 
  & 12.26561 & \multirow{2}{*}{17.92503 y} & 13.01911 & \multirow{2}{*}{119.80085 y} \\
  & $5/2^+ \rightarrow 3/2^-$ & nu & 7.87(3) &  & 0.07223268 
  & 7.611267 & & 8.435861 & \\ \hline  %---2nd
  \multirow{2}{*}{ $^{205}$Tl $\rightarrow$ $^{205}$Pb }  & 
    $1/2^+ \rightarrow 5/2^-$ &  u & ---  & \multirow{2}{*}{---} & 0.00013751 
  & 10.69247 & \multirow{2}{*}{58.43007 d} & 11.44598 & \multirow{2}{*}{305.13646 d} \\
  & $1/2^+ \rightarrow 1/2^-$ & nu & --- &  & 0.02785158 
  & 5.147932 & & 5.865505 & \\  %---3rd
  \bottomrule
\end{tabular*}
\end{table*}

%----------------------------------------------------- Discussions here :

In Figs. \ref{fig:fig1}, \ref{fig:fig2} and \ref{fig:fig3} we show the calculated excitation energies for low-lying states of the related parent and daughter nuclei. It is seen that for the deformed heavy odd-mass nuclei in the rare-earth region, $^{163}$Dy, $^{163}$Ho, $^{187}$Re and $^{187}$Os, the corresponding ground states are reproduced by our PSM calculations with correct spin-parity assignments. Besides, the ground-state rotational bands for these nuclei are described reasonably. For the heavier near-spherical odd-mass $^{205}$Tl and $^{205}$Pb, both the ground-state spin-parity assignments and the low excitation energies of the first-excited state are well reproduced. This indicates that the many-body wave functions of related nuclei are well described by our PSM calculations. With these wave functions, the calculated Log$ft$ values for related allowed and first-forbidden nuclear transitions are shown in Figs. \ref{fig:fig4}, \ref{fig:fig5} and \ref{fig:fig6}, where calculations without and with the quenching factors are shown in red brace and green parenthesis, respectively (the quenching factor of allowed transition is adopted as $f_q(\text{GT})=0.75$ as in Ref. \cite{martinez1996}), and compared with available data from Refs. \cite{Jung_163Dy_PRL_1992, Bosch_187Re_PRL_1996}. 

For neutral $^{163}$Dy atom, as seen from Fig. \ref{fig:fig4}(b) and Table \ref{tab1}, $Q_{\beta_{\text{c}}}$ is negative so that the common $\beta_{\text{c}}$ decay cannot occur. While in stellar finite-temperature environments, $^{163}$Dy may be fully ionized and then $\beta_{\text{b}}$-decay channel to both $K$ and $L$ shells is open as seen from Fig. \ref{fig:fig4}(a) and Table \ref{tab1}, which makes $^{163}$Dy as one of the important $s$-process branching points. In this case, the related nuclear transition between ground states are allowed transition and the transition strength is measured to be very strong with small Log$ft$ value, which is well reproduced by our PSM calculations when no quenching factor $f_q(\text{GT}) $ is adopted. By analyzing the corresponding wave functions of ground states, we found that the $5/2^-$ ground state of $^{163}$Dy has relatively pure (about $90.3\%$) $\nu 5/2^-[523]$ configuration originating from $\nu 1h_{9/2}$ orbital, and the $7/2^-$ ground state of $^{163}$Ho has relatively strong mixing of $\pi 7/2^-[523]$ configuration (about $82.8\%$) and many other higher-order qp configurations (but all of them include $\pi 7/2^-[523]$ and $\pi 5/2^-[532]$ levels that originate from $\pi 1h_{11/2}$ orbital). The GT transition operator is expected to have large matrix element between $\nu 1h_{9/2}$ and $\pi 1h_{11/2}$ orbitals \cite{Suhonen_book}, which makes that our calculated transition is strong with small Log$ft \approx 5.12$ ($\approx 5.36$) value without (with) quenching, as seen from Fig. \ref{fig:fig4}(a). 

With our calculated $f^\ast$ value shown in Table \ref{tab2} and the above calculated Log$ft$ values for $^{163}$Dy, the $\beta_{\text{b}}$-decay half-life of $^{163}$Dy is derived to be about 66 days (118 days) without (with) quenching as seen from Table \ref{tab2}. This indicates that the measured $\beta_{\text{b}}$-decay half-life of $^{163}$Dy can be described within a factor of two (three) by calculations without (with) quenching.

For neutral $^{187}$Re atom, it is unstable for $\beta_{\text{c}}$ decay with very small $Q_{\beta_{\text{c}}}$ value so that only transition between ground states exists which is unique first-forbidden transition, as seen from Fig. \ref{fig:fig5}(b) and Table \ref{tab1}. When fully ionized, $\beta_{\text{b}}$-decay channel is open and transition to the $3/2^-$ low-lying state of $^{187}$Os becomes possible, which is non-unique first-forbidden transition, as seen from Fig. \ref{fig:fig5}(a) and Table \ref{tab1}. The wave function of the $5/2^+$ ground state of $^{187}$Re is found to has $\pi 5/2^+[402]$ from the $2d_{5/2}$ orbital as the main configuration (about $54\%$) and strong mixing of many other higher-order qp configurations. The wave functions of the $1/2^-$ and $3/2^-$ states of $^{187}$Os are found to be very similar so that they form the ground-state band. $\nu 1/2^-[510]$ from the $3p_{3/2}$ orbital is found to be their main configuration (about $60\%$) with strong mixing of many other higher-order qp configurations. For the $5/2^+ \rightarrow 1/2^-$ unique first-forbidden transition, all the terms in Eq. (\ref{eq.all_ME}) vanish except for the nuclear matrix element involving tensor operator of rank two, i.e., $z$ in Eq. (\ref{eq.all_ME}d). The single-particle matrix element of such a transition operator is found to be small between $\pi 5/2^+[402]$ and $\nu 1/2^-[510]$ states, which lead to that our calculated transition strength is small with large Log$ft \approx 12.27$ without quenching by which the data can be described reasonably, as seen from Fig. \ref{fig:fig5}(a). For the $5/2^+ \rightarrow 3/2^-$ non-unique first-forbidden transition, all the terms in Eq. (\ref{eq.all_ME}) contribute in a complicated way analytically and numerically \cite{BLWang_1stF_2024}, the calculated transition strength is found to be larger with smaller Log$ft \approx 7.61$ without quenching, which can well reproduce the corresponding data as seen from Fig. \ref{fig:fig5}(a). With the calculated Log$ft$ and $f^\ast$ values, the derived $\beta_{\text{b}}$-decay half-life of $^{187}$Re is shown in Table \ref{tab2}, where it is seen that the measured $\beta_{\text{b}}$-decay half-life of $^{187}$Re can be described within a factor of two (four) by calculations without (with) quenching. 

Finally for the near-spherical $^{205}$Tl with small oblate deformation, neutral $^{205}$Tl atom is stable to $\beta_{\text{c}}$ decay with negative $Q_{\beta_{\text{c}}}$, while fully ionized $^{205}$Tl$^{81+}$ becomes unstable to $\beta_{\text{b}}$ decay where transitions from its ground state to the ground and very low-lying states of $^{205}$Pb are available as seen from Fig. \ref{fig:fig6}. This makes $^{205}$Tl as the last $s$-process branching point that decays to $^{205}$Pb which is $s$-only nucleus and is potential chronometers of the last $s$-process events and becomes probe of the corresponding stellar conditions. The $1/2^+$ ground state of $^{205}$Tl is found to have single-particle (qp) state of the $\pi 3s_{1/2}$ orbital as the main configuration (about $67\%$) of its wave function with strong mixing of many multi-qp configurations. The wave function of the $5/2^-$ ground state of $^{205}$Pb is found to has single-particle state from the $\nu 2f_{5/2}$ orbital as the main configuration (about $78\%$) and strong mixing of multi-qp configurations while the one of the $1/2^-$ low-lying state is pure to has single-particle state from the $\nu 3p_{1/2}$ orbital as the main configuration (about $94\%$). The first-forbidden transition to the $5/2^-$ ground state ($1/2^-$ low-lying state) is predicted to be weak (strong) with large (small) Log$ft$ value as seen from Fig. \ref{fig:fig6}. The $\beta_{\text{b}}$-decay half-life of $^{205}$Tl is predicted to be about 58 days and 305 days depending on if quenching factors are adopted or not, as shown in Table \ref{tab2}. 

In summary, bound-state $\beta$ decay is a crucial decay mode for nuclei in stellar environments, especially for the case of $s$ process. We provide a theoretical method to calculate and predict the bound-state $\beta$-decay half-lives of highly-ionized atoms so that the bound-state $\beta$-decay half-lives of nuclei from light to heavy ones including odd-mass and even-mass cases can be described systematically in a microscopic way for the first time. The method is based on our projected shell model that is developed very recently to take into account both allowed and first-forbidden transitions of nuclear $\beta$ decay, and combined with the traditional Takahashi-Yokoi model. We take three examples that are of much experimental interests, $^{163}$Dy$^{66+}$, $^{187}$Re$^{75+}$ and $^{205}$Tl$^{81+}$ as the examples for the first application. The ground states, low-lying states and available allowed and forbidden transition strengths of related nuclei are found to be described reasonably by our calculations. The bound-state $\beta$-decay half-lives of $^{163}$Dy$^{66+}$ and $^{187}$Re$^{75+}$ are described within a factor of two (four) when the corresponding nuclear matrix elements are not quenched (quenched) in our calculations. Besides, we predict that the bound-state $\beta$-decay half-life of the last $s$-process branching point $^{205}$Tl$^{81+}$ is about 58 and 305 days for cases without and with the quenching factors in calculations.

Near-future works include combining the method presented in this work and the Saha equation that is useful to determine the population of variously ionized states of heavy elements in realistic stellar conditions, after which it then would be possible for us to update the data table of $s$-process effective stellar $\beta$-decay rates \cite{TY_table_1987}.

%\nolinenumbers
\section*{Acknowledgements}
L.J.W. would like to thank B. H. Sun for discussions about bound-state $\beta$ decays. This work is supported by the National Natural Science Foundation of China (Grants No. 12275225), by the Fundamental Research Funds for the Central Universities (Grant No. SWUKT22050), and partially supported by the Key Laboratory of Nuclear Data (China Institute of Atomic Energy). 

\section*{Declaration of competing interest}
The authors declare that they have no known competing financial interests or personal relationships that could have appeared to influence the work reported in this paper.

\bibliographystyle{elsarticle-num}
%\bibliographystyle{cas-model2-names}

\bibliography{Refs_bound_state}

\end{document}